\documentclass[12pt]{article}
\usepackage{amssymb}
\newcommand{\Purple}[1]{#1}
\newcommand{\Red}[1]{#1}
\newcommand{\Blue}[1]{#1}
\newcommand{\ft}[2]{{\textstyle\frac{#1}{#2}}}
\newsavebox{\uuunit}
\sbox{\uuunit}
    {\setlength{\unitlength}{0.825em}
     \begin{picture}(0.6,0.7)
        \thinlines
        \put(0,0){\line(1,0){0.5}}
        \put(0.15,0){\line(0,1){0.7}}
        \put(0.35,0){\line(0,1){0.8}}
       \multiput(0.3,0.8)(-0.04,-0.02){12}{\rule{0.5pt}{0.5pt}}
     \end {picture}}
\newcommand {\unity}{\mathord{\!\usebox{\uuunit}}}
\def\rmi{{\rm i}}
\def\rmd{{\rm d}}

\newcommand{\largepart}[1]{\addtocounter{part}{1}
                      \vglue 0.7cm
                      {\Large \sc\noindent PART \thepart . #1}
                      \vglue 0.2cm
                      \addcontentsline{toc}{part}{\thepart.#1}
              }

\newcommand{\muv}{{\underline{\mu}}}
\newcommand{\nuv}{{\underline{\nu}}}
\newcommand{\rhov}{{\underline{\rho}}}
\newcommand{\sigmav}{{\underline{\sigma}}}
\newcommand{\dsl}{\not\!\partial }

\csname @addtoreset\endcsname{equation}{section}

\begin{document}

\begin{titlepage}
\begin{flushright}
UG-00-09\\ CERN-TH/2000-192 \\ KUL-TF-2000/20\\
hep-th/0007044
\end{flushright}
\vspace{.5cm}
\begin{center}
\baselineskip=16pt {\LARGE \bf
Supersymmetry in Singular Spaces }

\vfill
{\large Eric Bergshoeff $^{1}$, Renata Kallosh,$^{2,\dagger}$, 
and Antoine Van Proeyen $^{3,\ddag}$  } \\
\vfill
{\small
$^1$ Institute for Theoretical Physics, Nijenborgh 4,
9747 AG Groningen, The Netherlands \\[3mm]
$^2$ Theory Division, CERN, CH-1211 Gen\`{e}ve 23, Switzerland
\\[3mm]
${}^3$ Instituut voor Theoretische Fysica, Katholieke
 Universiteit Leuven,\\
Celestijnenlaan 200D B-3001 Leuven, Belgium
}
\end{center}
\vfill
\begin{center}
{\bf Abstract}
\end{center}
{\small We develop {\it the concept of supersymmetry in singular spaces},
apply it in an example for 3-branes in $D=5$ and comment on 8-branes in
$D=10$. The new construction has an interpretation that the brane is a
sink for the flux and requires adding to the standard supergravity a
$(D-1)$-form field and a supersymmetry singlet field. This allows a consistent
definition of supersymmetry on a ${S_1/ \mathbb{Z}_2}$ orbifold, the bulk
and the brane actions being separately supersymmetric.

Randall--Sundrum  brane-worlds can be reproduced in this framework
without fine tuning. For fixed scalars, the doubling of unbroken
supersymmetries takes place and the negative tension brane can be pushed
to infinity. In more general BPS domain walls with 1/2 of unbroken
supersymmetries, the distance between branes in some cases may be
restricted by the collapsing cycles of the Calabi--Yau manifold.

The energy of any static $x^5$-dependent bosonic configuration vanishes,
$E=0$, in analogy with the vanishing of the Hamiltonian in a closed universe.
}\vspace{2mm} \vfill \hrule width 3.cm {\footnotesize \noindent
$^\dagger$ On leave of
absence from Stanford University until 1 September 2000\\
 \noindent $^\ddag$ Onderzoeksdirecteur, FWO, Belgium }
\end{titlepage}
\tableofcontents{}

\section{Introduction}

Usually, supersymmetry is associated with non-negative energy. In
asymptotically flat spaces this is related to the time translation
operator which is a square of fermionic operators $E= Q^2 \geq 0$. In
curved space the positivity of energy theorem is proved via
Nestor--Israel--Witten construction where it is usually assumed that the
space is non-singular. In case of black holes with the singularity
covered by the horizon, the argument about the positivity of energy was
also extended \cite{GHHP}.

Recently there were number of reasons to reconsider the issues of
supersymmetry, in general. In the brane world scenarios with fine tuning
where singular branes are introduced \cite{RSI,RSII}, the status of the
supersymmetric embedding was not clear. The first attempts to relax the
fine tuning  was to find smooth solutions of supergravity where domain
walls are build from some scalar fields \cite{BCI,ST,ChG,DFGK,KLS}. With
respect to the `alternative to compactification scenario' \cite{RSII},
no-go theorems were established \cite{KL,BCII} for the BPS smooth
solutions of a certain class of supergravity theories ($N=2$ with vector
\cite{GST84,GST85} and tensor multiplets \cite{GunZagGaugMET}) on the
basis of the negative definiteness of the derivative of the
$\beta$-function in the relevant renormalization group behavior (UV   behavior).

More
recently, the complete $N=2$ supergravity theory was constructed in
\cite{AnnaGianguido} where also the hypermultiplets are included. The
status of hypermultiplets in the adS brane world is not yet
settled\footnote{J. Louis, talk at
SUSY2K, July 2000.}, more work is required. Even if the solutions with the IR fixed point will
be found, there will still be a problem to find a configuration in a
smooth supergravity which will connect two such IR fixed points. This may
be also explained following a suggestive argument\footnote{K. Stelle,
talks at Fradkin and G\"{u}rsey conferences, June 2000.}: BPS solutions are
expected to be given by harmonic functions, in codimension~1 they must
have a kink. A kink can not appear as a smooth solution. The
complimentary argument was given in \cite{GL}: the fermion mass does not
change sign when passing through the wall, which may explain the absence
of smooth solutions with the required properties. The version of a no-go theorem for
smooth brane worlds from compactifications, under some assumptions about
the potential, was recently proposed in \cite{malda}. None of these
arguments seems to give an unconditional final proof that there are no
smooth domain walls for a brane world scenario. However, they suggest to
find out whether the clear framework is available for the non-smooth
supersymmetric solutions.

Thus the purpose of this paper is to generalize supersymmetry for singular
spaces, where the curvature may have some $\delta$-function
singularities. In singular spaces there was no consistent and complete
definition of supersymmetry so far\footnote{We will comment on the
available information
\cite{HoravaWitten,Kellyandco,Bagger,Pomarol,Tomas,Polish} below.} and
the related issue of non-negative energy was not clearly addressed.

An important {\it reason to reformulate supersymmetry in singular spaces
is related to supersymmetric domain walls}: objects of codimension $d=1$,
i.e.~$(D-2)$-branes in $D$-dimensional space. They may be associated with
$D$-form fluxes which are dual to a scalar and piecewise constant. Such
forms do not fall off with  distance. In an infinite volume, they may
lead to an infinite energy and would be unphysical. Therefore, objects
like the $8$-brane in $D=10$ and 3-branes in $D=5$ may not exist as
independent objects. They may need some planes that serve as sinks for
the fluxes for supersymmetric configurations of codimension~1
\cite{PolBook}. Note that the fluxes in the branes of higher codimension
do vanish at infinity and this problem can be avoided.

Usually local supersymmetry is realized in supergravity when the
Lagrangian is integrated over a continuous space. Under supersymmetry, the
Lagrangian transforms as a total derivative. The parameters of local
supersymmetry are assumed to fall off at infinity and therefore the
action is invariant. If in addition to the supergravity action one
considers the $\kappa$-symmetric worldvolume action of the {\it positive
tension brane} of  codimension $d\geq 3$, it provides the
$\delta$-functional sources for the harmonic functions that describe the
configuration, $\partial_i\partial_i H= T \delta ^d(\vec y)$.

In the case of branes of  codimension $d=1$ defined on an orbifold ${S^1/
\mathbb{Z}_2}$, the harmonic functions satisfy the equations $\partial_y
\partial_y H= T \delta ( y)- T\delta(y-\tilde y)$. This can be seen e.g.
in the Ho\u{r}ava--Witten (HW) construction \cite{HoravaWitten} developed
for 5-dimensional 3-branes in \cite{Kellyandco}. The metric of such
solutions depends on $H= c+ T|y|$ and has two kinks at the orbifold fixed
points: one at $y=0$ and the other at $y=\tilde y$ where $\tilde y$ is
identified with $-\tilde y$. The supersymmetry in HW construction
includes the contribution from anomalies and requires quantum
consistency. It has some problematic features related to higher order
corrections and quartic fermionic terms.

To set up a new general point of view on supersymmetry in singular
spaces, and  to clarify the issue of energy of brane worlds, we introduce
here a set of rules defining consistent supersymmetry on singular spaces.

The most important new steps include: \textit{i)} some constants (masses
or gauge couplings) have to become ``$\mathbb{Z}_2$-odd'' to make
supersymmetry commuting with $\mathbb{Z}_2$-symmetry. This can be
achieved by promoting such a constant to the status  of a supersymmetry
singlet field, as suggested in \cite{EricAndFriends}; \textit{ii)}
moreover, one has to add to the theory  the $(D-1)$-form
potential\footnote{The importance of the $(D-1)$-form field was realized
by Duff and van Nieuwenhuizen \cite{Duff:1980qv}, who pointed out 20 years
ago the quantum inequivalence of the theories with and without such field
in the context of trace anomalies and 1-loop counterterms in topologically
non-trivial backgrounds. At about the same time Aurilia, Nicolai and
Townsend \cite{Aurilia:1980xj} have found that the $(D-1)$-form, which
propagates no physical particles, carries a surprising physics. They have
looked at $\theta$ parameter in QCD and at the cosmological constant in
supergravity.}  to compensate the variation of the Lagrangian
proportional to the derivative of the new field; \textit{iii)} one can
find afterwards the new bulk and the brane actions, which are separately
invariant under supersymmetry.

The {\it unusual features of the new supersymmetry} are the presence of
the supersymmetry singlet field and  $(D-1)$-form field in the bulk, and
the fact that the purely bosonic action on the brane is supersymmetric
due to the fact that its fermionic partner is a  $\mathbb{Z}_2$-odd
fermion which vanishes on the brane.

{\it In absence of brane actions, the new fields become irrelevant: on
shell for the 4-form and supersymmetry singlet, the bulk action reduces
to the standard supergravity action supersymmetric under the standard
rules}.

In general, when the brane actions are added, the new fields play an
important role in understanding the energy issue. We will find that {\it
the total energy of  supersymmetric configurations vanishes locally at
each brane.  The positive (negative) energy of the  brane tensions are
compensated  separately by the terms with the derivative of the
supersymmetry singlet field.} The energy of any static $y$-dependent
bosonic configuration vanishes locally, $E=0$, in analogy with the
vanishing of the Hamiltonian in a closed universe\footnote{We are
grateful to A. Linde who suggested this analogy.}.

The strategy is  applied  in detail in the particular case of a 3-brane in
$D=5$ on the basis of $U(1)$ gauged $N=2$ supergravity interacting with
abelian vector multiplets. We expect that it will work in other cases as
well: in $D=5$ one can try to include more general gauging and tensor and
hypermultiplets. The gauged $N=4$ and $N=8$ supergravities in $D=5$ are
also  natural candidates for an analogous extension. We will give a brief
discussion of the particularly interesting case of the 8-brane in $D=10$.

The paper has two main parts.

In \textsc{Part}~I, \textsc{The Supersymmetric Theory: Bosons And
Fermions}, we construct the supersymmetric actions in the bulk and in the
brane. To do so, we first identify the $\mathbb{Z}_2$ operator in
section~\ref{ss:susyZ2}. Then, we construct the supersymmetric theory
with three steps in section~\ref{ss:constrSUSYsing}. The final result is
written down in section~\ref{ss:sumd5S1Z2}

In \textsc{Part}~II, \textsc{The Background: Vanishing Fermions, Bosons
Solve Equation of Motion}, we study bosonic solutions of the theory from
part I and their unbroken supersymmetries. Section~\ref{ss:E0} discusses
the vanishing of the energy. The BPS equations and preserved
supersymmetries in these singular spaces are discussed in
section~\ref{ss:BPSsingsp}. We consider two cases, the fixed scalars,
with doubling of supersymmetry, and the general stabilization equations in
very special geometry, with 1/2 preserved supersymmetry. We show the
resulting formulas for the example of the STU wall, and for a particular
Calabi--Yau wall in section~\ref{ss:examples}.

Finally, in \textsc{Part}~III, we discuss the similar mechanism for the
8-brane in 10 dimensions.

The notations and use of indices are presented in
appendix~\ref{app:notations}. For those unfamiliar with the 5-dimensional
supergravity, and for establishing the related notations, we present its
structure in appendix~\ref{app:5dSGVM}. In appendix~\ref{app:previous} we
discuss the previous attempts to define supersymmetry on orbifolds.

\

\

\largepart{The Supersymmetric Theory:  Bosons And Fermions}

\section{Local supersymmetry and $\mathbb{Z}_2$}
\label{ss:susyZ2}
Our setup has the following basic features.
\begin{enumerate}
  \item We start with the standard  action $S_{bulk}$ of five-dimensional gauged
  supergravity with
   a symplectic Majorana  supersymmetry with parameters $\epsilon ^i$, $i=1,2$,
   thus having 8 real components,  coupled with
  an arbitrary number of vector multiplets \cite{GST84,GST85}.  
   \item \emph{Orbifold construction}. Fields live on a circle in the direction
  $x^5$, with an  orbifold condition. The circle implies that $\Phi (x^5)=\Phi
  (x^5 +2\tilde x^5)$, where $\tilde x^5$ is some arbitrary parameter
  setting the length of the circle. We use a concept of
  parity to split the fields in even and odd   under a $\mathbb{Z}_2$:
\begin{equation}
  \Phi _{even}(-x^5)=\Phi _{even}(x^5)\,,\qquad \Phi _{odd}(-x^5)=-\Phi _{odd}(x^5)\,.
\label{Phievenodd}
\end{equation}
This implies that the odd fields vanish at $x^5=0$ and at $x^5=\tilde
x^5$, where we will put the branes. The supersymmetries are also split in
half even ($\epsilon _+$) and half are odd ($\epsilon _-$). Both
$\epsilon _{\pm }$ have 4 real components. The bulk action is even, and
all transformation rules are consistent with the assignments.
\item \emph{The brane action} $S_{brane}$ is introduced. We place two branes, one at
$x^5=0$ and another one at $x_5=\tilde x_5$. The actions depend on the
values of the bulk fields at $x^5=0$ and $x_5=\tilde x_5$. As the odd
fields are zero on the brane, the brane action only depends on the even
fields. Only the supersymmetries $\epsilon _+$ act on these fields.
\end{enumerate}
We analyse all possibilities to make parity assignments consistent with
supersymmetry commuting with $\mathbb{Z}_2$-symmetry. We conclude that
the consistent supersymmetry on orbifolds without brane actions does not
fix the $\mathbb{Z}_2$-properties of the gauge coupling. However, after
adding the brane action we must require  the gauge coupling to be
$\mathbb{Z}_2$-odd\footnote{The second possibility, $\mathbb{Z}_2$-even
gauge coupling, as chosen in \cite{Bagger}, will be discussed in
Appendix~\ref{app:previous}.}.

We have to  treat the fact that $g$ is {\it only piecewise constant}. This
approach is inspired by \cite{EricAndFriends}. It consists of the
following steps
\begin{enumerate}
  \item Replace in the action the constant $g$ by a scalar function
  $G(x)$ and keep this field a {\it supersymmetry singlet}. The
  supersymmetry singlet field was introduced in the context of $\kappa$-symmetric
  brane actions in \cite{susysinglets}. This field has some
 peculiar properties: its shift under translation vanishes on shell but
 not off shell.
  \item Add a Lagrange multiplier $A_{\mu \nu \rho \sigma }$ that imposes the constraint $\partial
_\mu G=0$. At this point this reproduces the known 'bulk' actions of 5
dimensions \cite{GST84,GST85,GunZagGaugMET,AnnaGianguido}. The $G$-field
dependent action is invariant by appropriately choosing supersymmetry
transformations as well as local $U(1)$ gauge transformations for $A_{\mu
\nu \rho \sigma}$.
  \item Add source terms that are separately invariant, but may contain $\delta
  (x^5)$ and $\delta(x^5-\tilde
x^5)$ and are dependent on $A_{\muv\nuv\rhov\sigmav}$, where these are
the components in the 4 dimensions of the brane. After these additions the
  $A_{\muv\nuv\rhov\sigmav}$ field equation says that $\partial _5G(x)\propto \delta
  (x^5)- \delta(x^5-\tilde
x^5)$, and thus $G(x)\propto \varepsilon (x^5)$. This provides a
supersymmetric mechanism to change the sign of the coupling constant (and
of the fermion mass) when passing through the wall.
\end{enumerate}

We will first consider the possibilities for a $\mathbb{Z}_2$ in the
5-dimensional action. Define the operator $\Pi $ for any field, being
$+1$ or $-1$ whether it is even or odd under $x^5\rightarrow -x^5$. For
the symplectic Majorana spinors, the splitting might involve some
projection operators. Let us look for a $\mathbb{Z}_2$ of the form (for
any symplectic Majorana spinor $\lambda ^i$)
\begin{equation}
  \lambda ^i(x^5)=  \Pi (\lambda )\gamma _5 M^i{}_j \lambda ^j(-x^5)\,,
\label{projFerm}
\end{equation}
where $\Pi (\lambda )$ is thus a number $+1$ or $-1$, while $M^i{}_j$ is
so far an undetermined $2\times 2$ matrix. If this has to be a
$\mathbb{Z}_2$, the operation should square to $\unity $. Therefore $M$
should square to $\unity $. Notice that this is independent of whether we
included $\gamma _5$ in (\ref{projFerm}) or not. Next, this has to be
consistent with the reality condition (symplectic Majorana condition, see
appendix~\ref{app:notations}). This implies that $M$ should satisfy
$M^C\equiv \sigma _2 M^*\sigma _2=-M$, or thus, $M=\rmi m_0 \unity + m_1
\sigma _1 + m_2 \sigma _2 + m_3 \sigma _3$ with $m_0$, $m_1$, $m_2$ and
$m_3$ real. The $M^2=\unity $ condition implies that
\begin{equation}
  M^i{}_j=m_1 (\sigma _1)^i{}_j + m_2 (\sigma _2)^i{}_j + m_3 (\sigma
  _3)^i{}_j\,,
  \qquad \mbox{with}\ m_1,\, m_2,\,m_3\in \mathbb{R}\,,
\label{Mherm}
\end{equation}
which means that $M$ is hermitian and traceless. If $\gamma _5$ were not included
in (\ref{projFerm}), the numbers $m$ would have to be pure imaginary, and the
$M^2=\unity $ condition would imply $M=\unity $, but then we would have
no projection at all. Thus we conclude that (\ref{projFerm}) with
(\ref{Mherm}) is the only one that is possible.

Lowering the two indices on $M$, it becomes a symmetric matrix:
\begin{equation}
  M_{ij}=-m_1(\sigma _3)_{ij}-\rmi m_2(\unity )_{ij}+m_3(\sigma
  _1)_{ij}\,.
 \label{Mdown}
\end{equation}

The parity will be related to the fifth direction. We will therefore
assign a negative parity to the $x^5$ coordinate, or $\Pi (\partial
_5)=-1$. Consider now the supersymmetry transformation laws in the bulk,
see (\ref{susy5dV}). For one of the fermions we may arbitrary assign even
parity, e.g. for the components of the gravitino in the directions
excluding `5', i.e. $\psi _\muv $. Consider first the supersymmetry
transformation laws that are independent of the gauge coupling $g$. The
consistency of the parity assignments determines
\begin{eqnarray}
&& \Pi (e_\muv {}^m)=1 \,,\qquad \Pi ( e_5{}^5)=1\,,\qquad
\Pi ( A_5^I)=1\,,\nonumber\\
&& \Pi (e_5{}^m)=-1\,,\qquad \Pi
(e_\muv {}^5)=-1\,,\qquad \Pi (A^I_\muv)=\Pi(\Lambda^I)= -1\,, \nonumber\\
&&  \Pi (\psi _\muv )=\Pi (\epsilon )=1\,,\qquad \Pi (\psi _5)=\Pi
(\lambda)=-1\,.  \label{EvenOdd}
\end{eqnarray}
Note that also the supersymmetry parameters got a parity projection and
that the parameters of the gauge transformations, $\Lambda^I$, have to be
odd.

Now we consider the terms with $g$ in the gravitino transformation law
(\ref{susy5dV}). They depend on the constant matrix $Q_{ij}$, see
(\ref{Qreal}). Taking the parity transformation of both sides, we find
that they are proportional to
\begin{equation}
  M^j{}_iQ^k{}_{j}\epsilon _k=-\Pi (g) Q^j{}_i M^k{}_j\epsilon _k\,,
\label{MPPM}
\end{equation}
where we allowed a parity transformation of $g$. We find that if $Q$ and
$M$ commute (which means that they are proportional) one needs $\Pi
(g)=-1$, while if they anticommute (thus are taken in orthogonal
directions in the $SU(2)$-space), $\Pi (g)=1$.

Taking an anticommuting $M$ and $Q$ brings us to the setup of
\cite{Bagger}. We will see below, that the addition of brane actions
forces us to take a matrix that commutes with $Q$. If $Q$ and $M$ commute,
one has two possibilities to implement the parity assignment $\Pi (g)=-1$.
In the approach of \cite{Polish}, who take $M^i{}_j=(\sigma _3)^i{}_j$,
this assignment is realized by replacing $g$ by $g\varepsilon (x^5)$. On
the other hand, we will be able to make this assignment by promoting $g$
to a field.

To continue in the direction of \cite{Bagger,Pomarol,Polish}, one has to
provide the consistent definition of supersymmetry with step functions
and delta functions present in supersymmetry rules and in the action. The
construction of the higher order fermionic terms and the structure of the
algebra of supersymmetry in singular spaces in these approaches are
difficult as the HW   theory shows. Instead of this, a new way to
introduce supersymmetry in singular spaces will be developed below.

\section{Supersymmetry in a singular space}\label{ss:constrSUSYsing}
\subsection{Step 1: the bulk action}
We consider 5-dimensional supergravity coupled to $n$ vector multiplets.
The coupling is determined by a 3-index real symmetric constant tensor
$C_{IJK}$, where the indices run over $n+1$ values. We consider a $U(1)$
gauging of the $R$-symmetry group determined by real constants $V_I$. The
direction in $SU(2)$ space is determined by a matrix $Q_i{}^j$. Obviously,
the choice of that direction has no physical consequences. The full
action, to which we will refer as the GST action \cite{GST85}, and
transformation rules, are given in appendix~\ref{app:5dSGVM}. Replacing
the coupling constant by a scalar $G(x)$, the action is not any more
invariant supersymmetry,
\begin{equation}
e^{-1}  \delta(\epsilon) {\cal L}_{GST}= \left[ -\rmi \ft32\overline{\psi
}{}_\mu ^i\gamma ^{\mu \nu }\epsilon
  ^j W-\overline{\psi }{}_\mu^i\gamma ^{\mu \nu\rho  }\epsilon
  ^j A^{(R)}_\rho +\ft32\overline{\lambda }{}^i_x W^{,x}
  \gamma ^\nu \epsilon   ^j\right] Q_{ij}\partial _\nu G\,,
 \label{deltaSdgR}
\end{equation}
and neither under the $U(1)$ gauged symmetry,
\begin{equation}
e^{-1}  \delta_R {\cal L}_{GST}= \ft{1}{2}\left[
{\bar{\psi}}_{\mu}^{i}\gamma^{\mu\nu\rho}
   \psi^j _\rho
 +\overline{\lambda }{}^i_x\gamma ^\nu \lambda ^{jx}\right] Q _{ij}\Lambda _R
  \partial _\nu G\,,
 \label{gauge}
\end{equation}
where $\Lambda _R$ is the parameter of the $R$-symmetry
\begin{equation}
  \Lambda _R=V_I\Lambda ^I\,.
 \label{LambdaR}
\end{equation}

\subsection{Step 2: the four-form}
In the second step, we add the following Lagrange multiplier term:
\begin{equation}
  S_{A}=\frac{1}{4!}\int \rmd^5x\,\varepsilon ^{\mu \nu \rho \sigma\tau }A_{\mu \nu \rho \sigma }
  \partial _\tau G\,.
\label{SA}
\end{equation}
Now we can make the action invariant under supersymmetry. This is
obtained 1) by taking $G$ invariant under supersymmetry, and 2) by
defining the variation of $A$ such that all $\partial _\mu  G$ terms in
the transformation of the rest of the action are cancelled. The fact that
$G$ is invariant is consistent with the algebra because the translation of
$G$ is a field equation ($G$ is a supersymmetry singlet
\cite{susysinglets}). Thus the algebra is realized on-shell. The
resulting variation of $A_{\mu \nu \rho \sigma}$ under supersymmetry is
\begin{equation}
  \delta(\epsilon) A_{\mu \nu \rho \sigma }= \left[ -6\overline{\epsilon
  }^i \gamma _{[\mu \nu \rho }\psi _{\sigma ]}^jW+\rmi
  \overline{\epsilon }{}^i\gamma _{[\mu \nu }A_\rho ^{(R)}\psi _{\sigma
  ]}^j-\rmi\ft32\overline{\epsilon   }{}^i \gamma _{\mu \nu \rho \sigma }\lambda
  ^j_x W^{,x}\right] Q_{ij}\,.
 \label{deltaA}
\end{equation}
Under gauge transformations the 4-form transforms as follows
\begin{equation}
  \ft1{4!} \varepsilon ^{\mu \nu \rho \sigma\tau } \delta_R
   A_{\mu \nu \rho \sigma }=-\ft{1}{2}e\left[
{\bar{\psi}}_{\mu}^{i}\gamma^{\mu\tau\rho}
  \psi^j _\rho
+\overline{\lambda }{}^i_x\gamma ^\tau  \lambda ^{jx}\right]Q _{ij}
\Lambda_R\,.
 \label{gaugetrA}
\end{equation}
We define the covariant flux $\hat{F}$ as follows:
\begin{eqnarray}
\hat{F}&\equiv &\ft1{4!}e^{-1} \varepsilon ^{\mu \nu \rho \sigma \tau
}\partial _\mu A_{\nu \rho \sigma \tau } +\ft{1}{2}
{\bar{\psi}}_{\mu}^{i}\gamma^{\mu\nu\rho}A^{(R)}_\nu Q _{ij}\psi^j _\rho
 +\ft{1}{2}\overline{\lambda }{}^i_x\gamma ^\mu A^{(R)}_\mu Q _{ij}\lambda ^{jx}
\nonumber\\
&&+\ft32\left[\overline{\lambda }{}^i_x \gamma ^\mu \psi ^j_\mu W^{,x}
-\rmi\ft{1}{2}\overline{\psi}{}^i_\mu \gamma ^{\mu \nu } \psi _\nu ^j W
\right]Q_{ij}
 \,. \label{flux}
\end{eqnarray}
 The closure of the algebra on shell is due to a $G$ field equation
\begin{equation}
\hat{F} =
12 G \left (W^2- \frac{3}{4}({\partial W\over \partial
\varphi^x})^2\right)
+\rmi\overline{\lambda }{}^{ix}\lambda ^{jy} \left(-\ft14
g_{xy}W+\sqrt{\ft{3}{2}}T_{xyz}W^{,z}\right)Q_{ij}
 \,, \label{feG}
\end{equation}
where the bosonic part is related to the potential, given by
\begin{equation}
  V=-6 G^2 \left[W^2- \frac34({\partial W\over \partial
  \varphi^x})^2\right]\,.
 \label{pot_inW}
\end{equation}
This equation relates the flux to the potential via the singlet field $G$
(up to terms with fermions). We will see later that on shell the flux
will change sign when passing through the wall. This will explain the
role of the wall as a sink for the flux.

Remarks:\vspace{-3mm}
\begin{itemize}
  \item the action is invariant under 8 supersymmetries.
  \item the Lagrangian is invariant up to a total derivative. This is
  sufficient if the fields either drop off at infinity (as it is supposed
  to be in the 4-dimensional spacetime), or the space is cyclic and the fields
  are continuous (as it is supposed to be in the $x^5$ direction). To have
  at the end $G(x)$ piecewise constant, but not everywhere the same, it is
  clear that we need at least 2 branes where it can jump.
  \item The procedure outlined in step~2 is very general and does not
  depend on the details of the configuration.
\end{itemize}
We have explained in section~\ref{ss:susyZ2} that the GST-action allows
two different parity assignments for $G$. The action (\ref{SA}) respects
both choices with the assignments
\begin{equation}
 \left\{ \begin{array}{ccc}
 \Pi (G)=-1\,,&\qquad \Pi (A_{\muv\nuv\rhov\sigmav})=+1\,,&\qquad\Pi (A_{\muv
 \nuv\rhov 5})=-1\, , \\
 \Pi (G)=+1\,,&\qquad \Pi (A_{\muv\nuv\rhov\sigmav})=-1\,,&\qquad\Pi (A_{\muv
 \nuv\rhov 5})=+1\,.
 \end{array} \right.
\label{PiA5}
\end{equation}

\subsection{Step 3: the brane as a sink for the flux}
\label{ss:branesink}
 If we have only the bulk action, the field equation
still implies that $G$ is constant everywhere. The flux is proportional
to the potential and on shell for $A_{\mu\nu\rho\sigma\tau}$ and $G$ we
recover the standard supergravity.

We thus {\it need sources to modify the field equation on $G$}. These we
can choose according to a physical situation. For reproducing the
scenario described above, we take two branes positioned at $x_5=0$ and at
$x_5=\tilde x_5$. For both branes we introduce a worldvolume action,
which basically is the Dirac--Born--Infeld action in a curved background
with all excitations of the worldvolume fields set equal to zero.

The brane action includes the pullback of the metric, the scalars and the
4-form of the bulk action. The coefficient of the determinant of the
metric is taken to be the function $W$. It is related to the central
charge, similar to what was obtained for black holes in $N=2$, $d=4$ in
\cite{bhcy}. The Wess--Zumino term describes the charge of the domain
wall.

Remember that, as explained after (\ref{Phievenodd}), odd fields vanish
on the branes. Therefore, if we want to use the pullback of the components
$A_{\muv\nuv\rhov\sigmav}$ on the brane, we need that their parity is
even. This forces us to take the first choice in (\ref{PiA5}). This is
consistent with the scenario of \cite{Polish}, but it is problematic to
incorporate the approach of \cite{Bagger} in our framework of `consistent
supersymmetry on singular spaces'.

The brane action is
\begin{equation}
\Purple{  S_{brane}=-2 g \int d^5x\,\left( \delta (x^5)-\delta
(x^5-\tilde x^5 )\right)
  \left( e_{(4)} 3\alpha  W +\ft1{4!} \varepsilon ^{\muv \nuv \rhov \sigmav }A_{\muv \nuv \rhov
  \sigmav }\right)}\,,
\label{Sbrane}
\end{equation}
where $e_{(4)}$ is the determinant of the 4 by 4 vierbein $e_\muv^m$, and
$\varepsilon ^{\muv \nuv \rhov \sigmav}=\varepsilon ^{\muv \nuv \rhov
\sigmav 5} $ is in the same way a 4-density. The factor $\alpha =\pm 1$ is
a sign to be chosen later. The new field equation for $A_{\muv \nuv \rhov
\sigmav }$ is
\begin{equation}
  \partial _5 G(x^5)= 2 g\left( \delta (x^5)-\delta (x^5-\tilde x^5 )\right)\,,
\label{feGBrane}
\end{equation}
which has as solution
\begin{equation}
  G(x)=g\varepsilon (x^5)\,,
 \label{Gsolved}
\end{equation}
for $-\tilde x^5 <x^5<\tilde x^5 $. One should understand $\varepsilon
(x^5)$ as the function that is $+1$ for $0<x^5<\tilde x^5$, and $-1$ for
$-\tilde x^5<x^5<0$. Thus it has also a jump at $x^5=\tilde x^5\equiv
-\tilde x^5$, and
\begin{equation}
 \ft12 \partial _5\varepsilon (x^5)=\delta (x^5)-\delta (x^5-\tilde x^5)\,.
 \label{dvarepsx5}
\end{equation}
Now we may look at the equations of motion for the flux with account of
the value of the on-shell $G$-field. The bosonic part of the flux
(\ref{feG}) is on-shell
\begin{eqnarray}
\hat{F}_{on-shell,bos}= \varepsilon (x^5) 12 g \left [W^2-
\frac{3}{4}({\partial W\over
\partial \varphi^x})^2\right]
 \,. \label{feGonshell}
\end{eqnarray}
Clearly the flux is changing the sign when passing through the brane,
which justifies the title of this section. This may be contrasted with
the properties of the potential when the  $G$-field is on shell:
\begin{equation}
  V_{onshell}=  -6 g^2 \left[W^2- \frac34({\partial W\over \partial
  \varphi^x})^2\right]\,,
 \label{pot_inWg}
\end{equation}
where the standard assumption is made that $\varepsilon^2 (x^5)=1$. The
potential does not care about the existence of the brane.

The fermion mass terms  on shell for the $G$-field also change the sign
across the wall, as it follows from the terms in the action that are
quadratic in fermions and linear in $G$.

We now consider the invariance of the brane action. The f\"{u}nfbein satisfies
$e_5^m=e_\muv^5=0$. This is due to the parity assignment and orbifold
condition, which implies that only even parity fields are non-zero on the
brane. The variation of the brane action is ($A_\muv^{(R)}$ is zero on
the brane, and therefore those contributions can be neglected)
\begin{eqnarray}
  \delta S_{brane}&=&-3g\int \rmd^5x\left( \delta (x^5)-\delta (x^5-\tilde x^5 )\right)
 \nonumber\\ &&\times  \left[ W e_{(4)}   \bar \epsilon ^i\gamma ^m e_m^\muv \left(\alpha
  \psi _{\muv i} - \rmi \gamma _5 Q_{ij}\psi _\muv^j \right)+ W_{,x}
  \bar \epsilon  ^i\left( \rmi \alpha \lambda _i^x - \gamma _5 Q_{ij}
  \lambda ^{xj}
\right) \right]\,.
 \label{deltaSbr}
\end{eqnarray}
This vanishes if we apply the $\mathbb{Z}_2$ projections of
section~\ref{ss:susyZ2} with
\begin{equation}
  M_{ij}=\rmi \alpha Q_{ij}\,, \qquad \alpha=\pm 1\,.
 \label{MinQ}
\end{equation}
It thus implies that
\begin{eqnarray}
\psi _{\muv i}(x^5) &=& \rmi \alpha Q_{ij} \gamma _5\psi^j_{\muv
  }(-x^5)\,,
\nonumber\\
\lambda_i (x^5) &=& -\rmi \alpha Q_{ij} \gamma _5\lambda^j (-x^5)\,,
\nonumber\\
\epsilon _{ i}(x^5) &=&\rmi \alpha  Q_{ij}\gamma _5\epsilon^j(-x^5)\,,
\label{parfermions}
\end{eqnarray}
such that (\ref{deltaSbr}) vanishes.

\section{Summary of $d=5$ supersymmetry on $S^1/\mathbb{Z}_2$}
\label{ss:sumd5S1Z2}

In summary, the new Lagrangian in a singular space with the new
supersymmetry is given by
\begin{equation}
  S_{new}(x^5) =  S_{bulk}  + S_{brane}\,,\qquad  S_{bulk}  =
  S_{GST}(g\rightarrow G(x))- \int \rmd^5x\,e\, F(x)
 G(x)\,.
\label{newlagr}
\end{equation}
Here  $F$ is a curl of the 4-form $F=\frac{1}{4!} \varepsilon ^{\mu \nu
\rho \sigma\tau }\partial _\tau   A_{\mu \nu \rho \sigma }$ and
$S_{GST}(g)=S_0 +g S_1 + g^2 S_2$ is the standard $U(1)$ gauged
supergravity of \cite{GST85}, see appendix~\ref{app:5dSGVM}. The standard
supergravity at $g=0$ is called ungauged supergravity. We will use the
notation ${\cal L}_{0}$ for its Lagrangian.  Our new bulk theory has
Lagrangian ${\cal L}_{0}$, there are terms linear in $ G(x)$, which are
proportional to the flux $\hat F$ defined in  (\ref{flux}), and terms
quadratic in the field $G(x)$, see (\ref{pot_inW}),
\begin{eqnarray}\label{LagrangeVectorBulk}
\mathcal{L}_{bulk}&=& \mathcal{L}_{0} + 6 e  \Red{G^2 (x)}\left(
W^2-{3\over 4} W_{,x}^2\right) - e \,G(x)\hat F
 \nonumber\\
 && + e\;  \rmi\Red{G
(x)} Q_{ij} \overline{\lambda }{}^{ix}\lambda ^{jy} \left( -\ft{1}{4}
g_{xy}W +\sqrt{3\over 2}T_{xyz}W^{,z}\right)\, .
\end{eqnarray}

In addition, we will denote by $\delta_{0}$ the part of supersymmetry that
acts at $g=0$ and which forms the supersymmetry transformations of
ungauged supergravity. For completeness we repeat here the brane actions.
\begin{equation}
\Purple{  \mathcal{L}_{brane}=-2 g \,\left( \delta (x^5)-\delta
(x^5-\tilde x^5 )\right)
  \left( e_{(4)} 3 W +\ft1{4!} \varepsilon ^{\muv \nuv \rhov \sigmav }A_{\muv \nuv \rhov
  \sigmav }\right)}\,,
\label{SbraneA}
\end{equation}
The new supersymmetry rules are
\begin{eqnarray}
\delta (\epsilon) \{e_\mu^m, A_\mu^I, \varphi^x\}  & =&\delta_{0}
\{e_\mu^m, A_\mu^I, \varphi^x\}\,,  \nonumber\\
\delta (\epsilon )\psi_{\mu i} &=& \delta_{0} (\epsilon )\psi_{\mu i} +
\Red{G(x)} A_\mu ^{(R)}Q_{ij}\epsilon ^j+\rmi\ft{1}{2} \Red{G (x)}
\, \gamma_\mu \epsilon^j W Q_{ij}\,, \nonumber\\
\delta (\epsilon)\lambda _i^x&=&\delta_{0} (\epsilon ) \lambda _i^x -
\ft32\Red{G (x)}
\;\epsilon ^j  W^{,x}Q_{ij}\,, \nonumber\\
 \Blue{\ft1{4!}\delta (\epsilon) A_{\mu
\nu \rho \sigma}&=&    \left[ -6\overline{\epsilon
  }^i \gamma _{[\mu \nu \rho }\psi _{\sigma ]}^jW+\rmi
  \overline{\epsilon }{}^i\gamma _{[\mu \nu }A_\rho ^{(R)}\psi _{\sigma
  ]}^j-\rmi\ft32\overline{\epsilon   }{}^i \gamma _{\mu \nu \rho \sigma }\lambda
  ^j_x W^{,x}\right] Q_{ij}}\,,
\nonumber\\
\Red{\delta (\epsilon) G & =& 0 }\,.\label{newsusy}
\end{eqnarray}
The new gauge $R$-symmetry transformations are
\begin{eqnarray}
 \delta _R A_\mu ^{(R)} & = & \partial _\mu \Lambda_R \,, \nonumber\\
 \delta _R \psi _{\mu }^i & = & \Red{G(x)} \Lambda_R Q^{ij}\psi _{\mu j}\,,\nonumber\\
 \delta _R \lambda ^{xi} & = & \Red{G(x)} \Lambda_R Q^{ij}\lambda ^x_{\mu j}\,,\nonumber\\
  \Blue{\varepsilon ^{\mu \nu \rho \sigma\tau } \delta_R
  A_{\mu \nu \rho \sigma }&=&-\ft{1}{2}
{\bar{\psi}}_{\mu}^{i}\gamma^{\mu\tau\rho}
  \Lambda_R Q _{ij}\psi^j _\rho
 -\ft{1}{2}\overline{\lambda }{}^i_x\gamma ^\tau \Lambda_R Q _{ij}\lambda
 ^{jx}}\,.
 \label{newR}
\end{eqnarray}

 The action is defined by an integral over a
 product space ${\bf M}$ of the 4-dimensional manifold  and an orbifold, ${\bf M}= {\bf M}_4\times {S^1\over \mathbb{Z}_2}$.
\begin{equation}
  S^{new}= \int _{\bf M}d^4 x dx^5 {\cal L}^{new}\, .
\label{newaction}
\end{equation}
The 4d manifold is non-compact and the 5th dimension is compact but has
no boundaries in $S^1/ \mathbb{Z}_2$. Therefore, all surface  terms in the
variation of the action vanish assuming as usual that the parameters of
supersymmetry decrease at infinities of the 4d space. The bulk and the
brane actions are separately invariant. The supersymmetry transformations
form an on-shell closed algebra.

\

\

\largepart{The Background: Vanishing Fermions, Bosons Solve Equation of
Motion}

\section{Vanishing energy}
\label{ss:E0}

It is known that the Hamiltonian of the spatially closed universe
vanishes since in absence of boundaries  it is given by a diffeomorphism
constraint \cite{DeWittHartleHawking}. The basic argument goes as
follows. In 4d space when the ansatz for the metric is taken in the form
$ds^2= -(N^2- N_i N^i) dt^2+ 2N_i dx^i dt + h_{ij} dx^i dx^j$ one finds
that the Hamiltonian of   constraint is $H\equiv {\partial S \over
\partial N} =0$. Here $H$ has the contribution both from the gravity and
from matter. Still one has to keep in mind that the definition of the
energy of the closed universe is rather subtle.

In our new supersymmetric theory we also face the problem of how to define
the energy, in general. In our case the space is not an asymptotically
flat or an anti-de Sitter space. Moreover, since our space is singular, we
can not easily apply the Nestor--Israel--Witten construction, which for
asymptotically flat or anti-de Sitter  spaces would predict a non-negative
energy.

Therefore, we would perform here only a partial analysis of the energy
issue for supersymmetric theories in singular spaces, which has a clear
conceptual basis. We hope, however, that a more general treatment of this
problem is possible.

Here we limit ourselves to configurations which depend only on $x^5$. For
such configurations the natural definition of the energy functional was
suggested and studied in \cite{ST,DFGK,KL}.   We are using the warped
metric in the form
\begin{equation}
ds^2= a^2(x^5) dx^\muv dx^\nuv \eta_{\muv\nuv} + (dx^5)^2  \,.
\label{metric}
\end{equation}

Starting with the new Lagrangian (\ref{newlagr}), we may present the
energy functional for static $x^5$-dependent bosonic configurations as
follows
\begin{eqnarray}
E(x^5)&=&-6a^2 a^{\prime 2} +\ft12 a^4(\varphi^{x\prime} )^2
+a^4V-\ft1{4!}\varepsilon ^{\mu \nu \rho \sigma 5 }A_{\mu \nu \rho \sigma
}  G' \nonumber\\ && + 2 g \left(  \delta (x^5)-\delta (x^5-\tilde x^5
)\right)
  \left( 3  a^4\alpha W  + \ft{1}{4!}
  \varepsilon ^{\muv \nuv \rhov \sigmav }A_{\muv \nuv \rhov
  \sigmav }\right)\, .
\end{eqnarray}
Here ${}'$ means  $\partial \over \partial x^5$.  This expression in turn
can be given in the BPS-type form closely related to \cite{ST,DFGK,KL} but
still different due to \textit{i)} the presence of the 4-form and the
supersymmetry singlet off shell, \textit{ii)} the presence of $\alpha=
\pm$, which comes from the choice of the $\mathbb{Z}_2$ action on the
fermions and is introduced in (\ref{MinQ}).
\begin{eqnarray}
E(x^5)_{BPS}&=& \frac12  a^4 \left \{ \left[  \varphi^{x\prime} -
3\alpha  G
 W^{,x}  \right]^2 - 12 [  {a'\over a} +\alpha  G W]^2\right \} + 3 \alpha  [
a^4 G W]' \nonumber\\
 && + \left[ 2 g \left(  \delta (x^5)-\delta (x^5-\tilde x^5
)\right)-G'\right]
  \left( 3 a^4\alpha W  + \ft{1}{4!}\varepsilon ^{\muv \nuv \rhov \sigmav }A_{\muv \nuv \rhov
  \sigmav }\right)\,.\label{BPS1}
\end{eqnarray}
The tension of the first brane, $T^1_{x^5=0}$, is equal to $ 6 g \alpha W(
x^5=0) $. The tension of the second brane, $T^2_{x^5=\tilde x^5}$,  is
given by $- 6 g \alpha W( x^5=\tilde x^5) $. Even if any of them is
negative, this causes no problem since we have a compensating
contribution to the energy on each brane. In presence of the
supersymmetry singlet field, there is an additional contribution to the
energy at each brane due to the gradient of the supersymmetry singlet
field $G$. The term with $G'$ cancels the tension contributions at each
brane separately since  $G'= 2 g \left(  \delta (x^5)-\delta (x^5-\tilde
x^5 )\right) $ due to the field equations for the 4-form. With account of
the $A$ and $G$ equations and no boundary condition, which allows to
ignore $+ 3 \alpha  [ a^4 G W]'$, the energy functional takes the form
\begin{eqnarray}
E(x^5)_{BPS}&=& \frac12  a^4 \left \{ \left[  \varphi^{x\prime} - 3\alpha
G
 W^{,x}  \right]^2 - 12 [  {a'\over a} +\alpha  G W]^2\right \}\,.
\end{eqnarray}
Note that the {\it energy functional is still not positive definite}: one
perfect square with the kinetic energy of the normal scalar $\varphi$ is
positive, however that with the kinetic energy of the conformal factor
of the metric is negative, as it should be. One might have a concern
about some configurations  where the negative contribution will dominate
over the positive one, which will lead to the instability of the theory.
However, using the  equations of motion  for $g_{00}$ and $A_{\mu \nu
\rho \sigma }$ (not the BPS equations), one finds that the energy vanishes
for any $x^5$-dependent solution of the equations of motion. This can
also be reinterpreted as derived using the canonical formulation of the
gravitational theory where one starts from
\begin{equation}
  \int _M -\ft12 \sqrt{g}R +\int _{\partial M}{\cal K}=\int \tilde {\cal L}\,,
 \label{gRK}
\end{equation}
where ${\cal K}$ is such that the resulting action has no second
derivatives on the metric. The action of matter fields can be added  and
as in \cite{DeWittHartleHawking} one finds that
 an on-shell energy $E$, at least for time-independent
configurations, vanishes
\begin{equation}
\oint E =0
 \label{cintE=0}
\end{equation}
for any solution of the equations of motion.

One very important property of the vanishing of the energy of the closed
universe is that it vanishes locally and not due to the compensation of
the total energy, positive and negative, in different parts of the
universe. As we explained this happens also here: the cancellation of
energy on each brane takes case separately: the tension term is cancelled
by the energy of the supersymmetry singlet field $G(x)$.

\section{BPS construction in singular spaces}\label{ss:BPSsingsp}

The squared terms in (\ref{BPS1}) suggest the BPS conditions:
\begin{equation}
g_{xy}\,(\varphi^y)'  =  + 3 \alpha  G(x^5)  W_{,x} \ , \qquad \qquad
{a'\over a} = - \alpha  G(x^5)  W \,, \label{BPSsol}
\end{equation}
where we can use the on shell value  of  $G(x^5)$, which is equal to $ g
\varepsilon (x^5)$, so that we obtain
\begin{equation}
g_{xy}\,(\varphi^y)'  =  + 3 \alpha g \varepsilon (x^5)  W_{,x} \ ,
\qquad \qquad {a'\over a} = -  \alpha g \varepsilon (x^5)  W \,.
\label{BPSsolonshell}
\end{equation}
{}From now  on we will make the choice   $\alpha =1$, i.e. pick up a
particular property of fermions under parity. Physics depends on the sign
of $\alpha g$. Therefore, the change in the sign of $\alpha$ can always be
compensated by a change of the sign of $g$.

One can verify that the jump conditions on the branes\footnote{It was
observed in \cite{BS} that in $N=8$ gauged supergravity the jump
conditions on the branes may be satisfied if the tensions are related to
the superpotential. However, putting the step functions in supersymmetry
transformations by hand may in general cause problems with higher order
corrections. Our work makes it plausible that $N=8$ gauged supergravity
with addition of the flux and supersymmetry singlet may be constructed
with the complete and consistent supersymmetry. This will generate the
step functions in supersymmetry rules in presence of branes. }, derived
starting with the second order differential equations, are satisfied
automatically due to the new supersymmetry (\ref{newsusy}), the on-shell
condition (\ref{feGBrane}) and the presence of the 5-form flux, changing
the sign when passing through the wall.

Let us also consider the Killing spinors. In the background with only
non-vanishing scalars and the warped metric (\ref{metric}), which has as
only non-zero components of the spin connection $\omega _\muv ^{m5}=
a'\delta_\muv ^m$, the transformations of the spinors are
\begin{eqnarray}
 \delta (\epsilon )\lambda _i^x & = & -\rmi\ft12 \gamma _5\varphi
 ^{x\prime}\epsilon _i
 -\ft32 G W^x Q_{ij}\epsilon ^j\, , \nonumber\\
 \delta (\epsilon )\psi _{\muv i} & = &\partial _\muv \epsilon _i
 + \ft12 \delta _\muv^m \gamma
 _m\left( a'\gamma _5\epsilon _i+\rmi a G W Q_{ij}\epsilon ^j\right)\, ,
  \nonumber\\
 \delta (\epsilon )\psi _{5i} &=& \epsilon '_i+\ft12\rmi G W\gamma _5Q_{ij}\epsilon ^j
 \,.
 \label{Killing1}
\end{eqnarray}
To solve these, we split
\begin{eqnarray}
 \epsilon _i & = &\epsilon _i^+ + \epsilon _i^-\,, \nonumber\\
 \epsilon _i^\pm & = & \ft12
 \left(\epsilon _i \pm \rmi \gamma _5Q_{ij}\epsilon ^j\right)=
 \pm \rmi\gamma _5Q_{ij}\epsilon ^{\pm j} \,.
 \label{eps+-}
\end{eqnarray}
Using the conditions (\ref{BPSsol}), the last equation of (\ref{Killing1})
gives the dependence of the supersymmetries on $x_5$. We obtain
\begin{equation}
  \epsilon _i^\pm =a^{\pm 1/2}\epsilon _i^\pm (x^\muv)\,.
 \label{eps5dep}
\end{equation}
The second equation gives
\begin{equation}
  \partial _\muv \epsilon _i^++\partial _\muv \epsilon _i^-+\delta _\muv^m \gamma
 _m a'\gamma _5\epsilon _i^-=0\,.
 \label{grav+-trans}
\end{equation}
The solutions are thus
\begin{equation}
  \epsilon _i= a^{1/2}\epsilon _i^{+(0)}
  +a^{-1/2}\left(1-\frac{a'}{a}x^\muv \gamma _\muv\gamma _5\right)
  \epsilon_i^{-(0)}\,,
 \label{2solKill}
\end{equation}
where $\epsilon_i^{\pm (0)}$ are constant spinors with each only 4 real
components due to the projection (\ref{eps+-}). Remains the first Killing
equation, which implies
\begin{equation}
  \varphi ^{x\prime}\epsilon_i^{-(0)}=0\,.
 \label{gauginoKilling}
\end{equation}
There are thus two possibilities to solve the Killing equations.
\begin{itemize}
\item Maximal unbroken supersymmetry in the bulk ($N=2$).

Here we require that the scalars are strictly constant, $\varphi
^{x\prime}=0$, and the superpotential is independent of the scalars,
${\partial W\over \partial \varphi^x}=0$, at the solution. No constraints
on Killing spinors arise from the gaugino.

{}From the gravitino transformation, we have shown in (\ref{2solKill})
that in the warped geometry a result similar to \cite{LPT} takes place,
with doubling of supersymmetries: the constant spinors $\epsilon_i^{\pm
(0)}$ give together 8 unbroken supersymmetries.

\item 1/2 of the maximal unbroken supersymmetry in the bulk ($N=1$).

When the scalars are not constant, then there is the extra condition
(\ref{gauginoKilling}), leaving just one projected supersymmetry with 4
real components,
\begin{equation}
  \epsilon _i -\rmi \gamma _5Q_{ij}\epsilon ^j=0\,, \qquad
  \epsilon _i= a^{1/2} \epsilon _i^{(0)}\,,
 \label{BPSepsilon}
\end{equation}
where $\epsilon _i^{(0)}$ is constant. Note that this projection of the
supersymmetries is on the brane consistent with (\ref{parfermions}).
Thus, we remain in the bulk as well as on the brane with $1/2$ of the
original supersymmetries. Vice versa, imposing the projection
(\ref{BPSepsilon}) one derives from the vanishing of (\ref{Killing1})
that the conditions (\ref{BPSsol}) should be satisfied.
\end{itemize}

\subsection{Fixed scalars, doubling of supersymmetries\\
 and an alternative to compactification world brane}

The field equation for the 4-form and  $G$-field Killing equations are
solved if (\ref{BPSsol}) are solved. In this section we look for the very
particular solutions of these equations with maximal unbroken
supersymmetry that have everywhere constant scalars\footnote{These
solutions remind the so called double-extreme black holes
\cite{double,critical}, which have fixed scalars in their solutions
\cite{FKS}.}.

The `fixed scalar domain wall solution' is given by
\begin{equation}
(\varphi^y)'  = 0 \ , \qquad
 \left({\partial W \over \partial \varphi^x}\right)_{crit} =0 \ ,
 \qquad \qquad \frac{a'}{a} = -  g \varepsilon (x^5)  W_{crit} \,.
\label{double}
\end{equation}
The  solution is given by the  supersymmetric attractor equation
\cite{FKS,critical} in the form
\begin{equation}\label{attractor}
  C_{IJK}\bar h^J \bar h^K  = q_I  \,,
\end{equation}
where
\begin{equation}
  \bar h^K \equiv  \sqrt{W_{crit}} h^K\,,
\label{barh}
\end{equation}
and we have used charges normalized as
\begin{equation}
  q_I\equiv \sqrt{\ft23} V_I\,, \qquad \mbox{such that}\qquad
  W= h^Iq_I\,,\qquad W_{,x}=-\sqrt{\ft23}h_x^Iq_I\,.
 \label{defq}
\end{equation}
Consistency implies
\begin{equation}
  W_{crit}(C_{IJK}, q_I)=h_{crit}^Iq_I= (\bar h^I q_I)^{2/3}= (C_{IJK}\bar
  h^I  \bar h^J \bar h^K)^{2/3}\,.
\label{Wcr}
\end{equation}
In some cases the explicit solution of the attractor equation is known in the form
\begin{equation}
\bar h^I(C_{IJK}, q_I)
\label{explicit}
\end{equation}
(see e.g. \cite{marina,critical} where many examples
are given). The metric is
 \begin{equation}
ds^2= e^{-2 g W_{crit}|x^5|} dx^\muv dx^\nuv \eta_{\muv\nuv} + (dx^5)^2
\,. \label{metric3}
\end{equation}

If we choose $gW_{crit}$ to be positive (which means that at $x^5=0$ we
have a positive tension brane), the metric is that introduced in
\cite{RSI}, where two branes are present at some finite distance from
each other. We will refer to this scenario as RSI. The second brane has a
negative tension and can be sent to infinity, in principle, which leads
to an alternative to compactification, discussed in \cite{RSII}. We will
refer to this as RSII. Whether this limit is totally consistent is an
independent issue, however the warp factor in the metric can be chosen to
exponentially decrease away from the positive tension brane. This is not
in a contradiction with the no-go theorem \cite{KL,BCII}. For constant
scalars at the critical point, the metric behaves differently from the
case when scalars are not constant but approaching the critical point.
This can also be explained by the {\it doubling of unbroken
supersymmetries in the bulk} for these solutions. Indeed, the gaugino
transformations are vanishing without any constraint on the Killing
spinors and the gravitino transformations also have an 8-dimensional zero
mode as shown in (\ref{2solKill}). Note that the curvature is everywhere
constant, except on the branes  where the metric has a cusp. For example,
the scalar curvature for this solution, is equal to
\begin{equation}
\ft14R= 5 g^2 W_{crit}^2 -  g (\delta(x^5) - \delta(x^5- \tilde
x^5))W_{crit}\,.
 \label{curv}
\end{equation}

The simplest example of fixed scalar domain wall (related to
double-extreme $STU$ black holes) comes out from the M5-brane compactified
on $T^6$ so that $C_{IJK}h^I h^J h^K= STU=1$ and $W_{crit}= 3 (q_S q_T
q_U)^{1/3}$ (see sec.~4.3  in \cite{critical}). Here $q_S, q_T, q_U$ are
FI terms.

\subsection{BPS equations in very special geometry (with vector multiplets)}
In the context of our present work in the more general case with vector
multiplets present and non-constant scalars, we will first change
coordinates, write down the energy functional in the new coordinate
system and proceed  by solving the BPS conditions following from the
energy functional. We take
\begin{equation}
ds^2= a^2(y) dx^\muv dx^\nuv \eta_{\muv\nuv} + a^{-4}(y)dy^2  \,,
\label{metrica}
\end{equation}
so that ${\partial\over \partial x^5}= a^2 {\partial \over \partial y}$
and we will use
 $\dot {}$ for $\partial \over \partial y$.
 The  BPS-type energy
functional for static $y$-dependent bosonic configurations following from
the new Lagrangian (\ref{newlagr}) is
\begin{eqnarray}
E(y)&=& {1\over 2}  a^2 \left \{ [  a^2 \dot {\varphi^x} - 3 G
 W ^{,x} ]^2 - 12 [  a \dot a + G W]^2\right \} + 3\frac{d}{dy}
[ a^4 G W]\label{BPS2} \nonumber\\
 & & + \left[ 2 g \left(  \delta (y^5)-\delta (y^5-\tilde y^5
)\right)-\dot G\right]
  \left( 3 a^4\alpha W  + \ft{1}{4!}\varepsilon ^{\muv \nuv \rhov \sigmav }A_{\muv \nuv \rhov
  \sigmav }\right)\,.\nonumber
\end{eqnarray}
The stabilization equations for the energy are the BPS conditions:
\begin{equation}\label{stab}
 a^2 g_{xy} \dot{ \varphi^y}  =  + 3  G(y)    W_{,x} \,, \qquad \qquad
a \dot a = -  G(y) W\,,
\end{equation}
where on shell for the 4-form field
\begin{equation}
  G(y) = g \varepsilon (y)
\ , \qquad  \dot G = 2g\left( \delta (y)-\delta (y-\tilde y )\right)
\label{Gy}\,.
\end{equation}
The solution of these closely related equations is known both in the
context of black holes \cite{sabra} and domain walls \cite{Klaus} in
smooth supergravities. The difference is that we have now derived the BPS
equations from supersymmetry with the step functions, and the signs are
correlated with the  choice of the parity assignments for fermions. This
takes care of the jump conditions on the branes where our solutions have
kinks.

The equations (\ref{stab}) can be combined to one $(n+1)$-component
equation with a free index $I$ in very special geometry. The equations
(\ref{defquantVM}) imply for any derivative
\begin{equation}
  g_{xy}\dot \varphi ^y= -\sqrt{\ft32}h_{Ix}\dot h^I= \sqrt{\ft32}h^I_x\dot
  h_I\,.
 \label{gxydphi}
\end{equation}
The BPS equations are then
\begin{eqnarray}
  h_x^I \left( a^2\dot h_I+2 q_IG(y)\right) =0\,,&\qquad & 2a\dot a
  +2 h^Iq_IG(y)=0\,, \nonumber\\
h_x^I \left( \frac{d}{dy}(a^2 h_I)+2 q_IG(y)\right) =0\,,&\qquad &
h^I\left( \frac{d}{dy}(a^2 h_I)  +2 q_IG(y)\right) =0\,.
 \label{BPSspgeomsplit}
\end{eqnarray}
These $n+1$ equations are equivalent to
\begin{equation}
   \frac{d}{dy}(a^2 h_I)+2 q_IG(y) =0\,.
 \label{BPSspgeom}
\end{equation}
We can rewrite it using the expression for the dual coordinate $h_I =
 C_{IJK}h^J h^K$
\begin{equation}\label{stab4}
   {d\over d y} ( C_{IJK}\tilde h^J \tilde h^K)  = - 2 G(y)q_I
    \qquad \mbox{where}\qquad  \tilde h^I \equiv a(y) h^I \,.
\end{equation}
\subsection{ Supersymmetric Domain Walls with non-constant scalars}
\label{ss:DWnonconstsc}

Using the on-shell expression for $G(y)$ we may solve this and get that
\begin{equation}\label{stab5}
  C_{IJK}\tilde h^J \tilde h^K  = H_I(y)   \,,
\end{equation}
where $H_I$ is a harmonic function,
\begin{equation}\label{stab6}
   H_I = c_I - 2g q_I |y|\,,
\end{equation}
satisfying the equation
\begin{equation}
 {d\over d y}  {d\over d y} H_I=
 -4g q_I [\delta(y) - \delta(y-\tilde y)]\,.
\label{complete}
\end{equation}
This is an explicit answer for a given $C_{IJK}$ and $H_I$ under the
condition that we know how to solve the algebraic attractor equation
(\ref{attractor}). If we know the solution of the algebraic attractor
equation (\ref{attractor}) in the form  (\ref{explicit}), we simply
replace the $q_I$ in this expression by the harmonic functions $H_I(y)$.
\begin{equation}
\tilde h^I(y) = \tilde h^I (C_{IJK}, H_I(y))\,, \label{solution}
\end{equation}
and we can find the scalars and the metric. Contracting the attractor
equation with $\tilde h^I$ we find that
\begin{equation}\label{stab55}
  C_{IJK}\tilde h^I \tilde h^J \tilde h^K  = a^3 (y)   C_{IJK} h^I  h^J h^K
   =  a^3 (y)\qquad \mbox{or}\qquad a^2(y)=h^IH_I\,.
\end{equation}
Our solution for the domain wall metric is therefore given by
(\ref{metrica}) with $a(y)$ given by this expression.

The choice of coordinates for the scalars $\varphi ^x$ has been left
unspecified so far. One possible choice is that similar to the
`special coordinates' in $d=4$, $N=2$:
\begin{equation}
  \phi^I = {\tilde h^I(y)\over \tilde h^0(y)}= \{
1, \varphi^x(y)\} \label{mod}\, .
\end{equation}

The general class of domain wall solutions as described in the previous
section has many particular realizations, depending on the choice of the
intersection numbers $C_{IJK}$ as well as on the choice of the initial
conditions at $y=0$ defined by the first terms in the harmonic functions,
$c_I$. Only if at least some of the $c_I$ are not vanishing, would the
scalars depend on $y$. Otherwise all ratios of harmonic functions will be
$y$-independent and therefore all scalars will be constants and the
solutions will be those from the previous section.

The properties of the solutions with non-constant scalars depend on the
choice of the model, i.e. the choice of the intersection numbers
$C_{IJK}$, and also on various choices of the signs between the first and
the second term in the harmonic functions. The options are:

\begin{itemize}
\item All $c_I$ have the opposite sign to all $g q_I$, i.e. none of the harmonic
functions vanishes at some value of $y$, $|H_I|= |c_I| +|2gq_I||y| >0$.
In such case one can directly use most of the solutions of the
stabilization equations from \cite{critical} and present the relevant
domain walls. Particularly interesting ones require the second wall to
cut off the part of the 5D space where the relevant CY cycles collapse
and/or the metric of the moduli space vanishes.

\item At least some of the  $c_I$ have the same sign as
$g q_I$, e.g. for positive $c_1$ and positive $gq_1$, $H_1= c_1
-2gq_1|y|$ vanishes at $|y|_{sing}={c_1\over 2gq_1}$. Such cases
definitely require the second brane at $|\tilde y|$ to be placed before
$|y|_{sing} $ since at  $|y|_{sing}$ the metric of the moduli space and
the space-time metric may vanish.
\end{itemize}

\section{Examples}\label{ss:examples}

Here we briefly present a couple of examples. It would be interesting to
undertake a more careful study of these solutions.

\subsection{STU Wall}
We consider again  $C_{IJK}h^I h^J h^K= STU=1$ and the harmonic functions
are given by
\begin{equation}
  H_S=c_S-2 g q_S |y|\,, \qquad   H_T=c_T-2 g q_T |y|\,, \qquad
  H_U=c_U-2 g q_U |y|\,.
\label{harmonicSTU}
\end{equation}
If all $c$ are positive and all $gq$ are negative, all harmonic functions
are strictly positive. The metric  is
\begin{equation}
  ds^2 =3(H_S H_T H_U)^{1/3}(dx)^2 + 3^{-2}(H_S H_T H_U)^{-2/3}dy^2\,,
\label{STUmetric}
\end{equation}
and the moduli are
\begin{equation}
  S= \left({H_T H_U \over H_S^2}\right)^{1/3}\,,\qquad   T= \left({H_S H_U \over
  H_T^2}\right)^{1/3}\,, \qquad   U= \left({H_T H_S \over
  H_U^2}\right)^{1/3}\,.
\label{moduli}
\end{equation}
We use the normalization $3 (c_S c_T c_U)^{1/3}=1$
 so that
near $|y|=0$ the domain wall metric (\ref{metrica}) tends to a flat
 Minkowski metric.
\begin{equation}
ds^2_{y \rightarrow 0}\Rightarrow    dx^\muv dx^\nuv \eta_{\muv\nuv}\ + d
y^2\,. \label{DWmetric2}
\end{equation}
If near the  second wall at $|y| = |\tilde y|$  the values of the second
terms in the harmonic functions are much larger than the first ones, the
metric near this brane approaches the boundary of the $adS$ space with
\begin{equation}
ds^2  \Rightarrow
 \rho^2 dx^\muv dx^\nuv \eta_{\muv\nuv}+  R_{AdS}^2 \left({d \rho\over
\rho}\right)^2 \, , \label{DWmetric22}
\end{equation}
where  $y= \ft12 R_{AdS}\rho^2$ and  $R^{-2}_{AdS}= 9 g^2 (q_S q_T
q_U)^{2/3}$.

If in any of the harmonic functions the sign between the terms is
opposite, and the harmonic function vanishes at $ |y|_{sing}$, the second
brane has to be at the position $|\tilde y|< |y|_{sing}$.

\subsection{Calabi-Yau Wall (Base ${\bf P}_2$ Vacuum)}

Here we take an example of a large complex structure limit of a
particular Calabi-Yau threefold \cite{Candelas} which defines a cubic form
for 5D supergravity.
\begin{equation}
{\cal V}(h^1,h^2)=9(h^1)^3+9 (h^1)^2 h^2+3 h^1 (h^2)^2=1\,.
\end{equation}
The attractor equation for this theory was solved in \cite{critical}. We
will use this solution to present one for the CY domain wall. We perform a
coordinate redefinition to variables $U$ and $T$, related to the basic
cycles $h^1$ and $h^2$ by
$U=6^{\frac{1}{3}}h^1$, $T=6^{\frac{1}{3}}(h^2+\frac{3}{2}h^1)$ and
\begin{equation}
{\cal V}(U,T)=\frac{3}{8}U^3+\frac{1}{2}U T^2=1\,. \label{precandTU}
\end{equation}
The relevant harmonic functions are given by
\begin{equation}
  H_U=c_U-2 g q_U |y|\,, \qquad   H_T=c_T-2 g q_T |y|\,,
\end{equation}
and we choose $H_U>0$ and $H_T>0$. The domain wall metric takes the form
(\ref{metrica}), where
\begin{equation}
a^2(|y|)= \left[H_U\frac{2}{\sqrt{3}}
\sqrt{H_U-\sqrt{H_U^2-\frac{9}{4}H_T^2}} +H_T \sqrt{3}\sqrt{ H_U+
\sqrt{H_U^2-\frac{9}{4}H_T^2}}\right]^{2/3} \,. \label{CYmetrica}
\end{equation}
The moduli are given by
\begin{eqnarray}
U(|y|)=\frac{2}{\sqrt{3} a(|y|)}
\sqrt{H_U-\sqrt{H_U^2-\frac{9}{4}H_T^2}}\,,\qquad
T(|y|)=\frac{\sqrt{3}}{a(|y|)} \sqrt{H_U+\sqrt{H_U^2-\frac{9}{4}H_T^2}}\,.
\label{stabcand}
\end{eqnarray}
The range of the variables is determined by the positivity of the basic
cycles $h^1$ and $h^2$, which translates for our solution into
\begin{equation}
2 H_U- 3 H_T > 0\, , \qquad H_U > 0 \, , \qquad H_T > 0 \, .\label{candrest}
\end{equation}
To satisfy this condition at $|y|=0$ we have to require that $c_U>0$,
$c_T>0$ and $2c_U-3c_T>0$. Note that if $H_T$  would vanish, which may
happen for $g q_T>0$ and $c_T>0$ at $|y|_{sing}={c_T\over 2gq_T}$, this
would be the point where the cycle $h^1$ would collapse. Indeed at
$H_T=0$, the ratio ${U\over T}= {h^1\over h^2 + (3/2) h^1} = 0$. At this
point also $a(|y|_{sing})=0$ and the spacetime metric is singular. To
avoid such a singularity we may require that the second brane is at
\begin{equation}
  |\tilde y|< |y|_{sing}={c_T\over 2gq_T}\,.
\label{collapse}
\end{equation}
Even if none of the harmonic functions $H_U$ and $H_T$ vanishes, i.e.
$gq_T$ and $gq_U$ are both negative, we still have an inequality
\begin{equation}
2 c_U -3 c_T - |y| 2g (2 q_U-3q_T)> 0\,. \label{inequality}
\end{equation}
If  the  condition  $g(2 q_U-3q_T)>0$ is satisfied, then there is also an
upper limit for the distance between the branes:
\begin{equation}
|\tilde y| < {2 c_U - 3 c_T \over 2g(  2 q_U- 3 q_T) }\,.
\end{equation}
At this point $h^2 =6^{-1/3}(T- \ft32 U)= 0$, i.e. the second cycle is
collapsing. Thus to prevent this from happening, we have to put the second
wall before this value of $|y|$.

As we already mentioned, a more careful and detailed analysis of  domain
walls of section~\ref{ss:DWnonconstsc} in application to other CY spaces,
as it was done for CY black holes, may lead to more interesting and
diverse configurations.

\

\

\largepart{Discussion:10D supersymmetry and 8-branes on the orbifold.}
\label{ss:8brane} \addtocounter{section}{1} \setcounter{equation}{0}

The massive 10D supergravity theory of Romans \cite{Romans} describing
the  8-brane is expected to have the same basic features as 5D
supergravity describing the 3-brane. Thus we expect using
\cite{Polchinski:1996df,EricAndFriends} and the results of the present
paper that the 10D supersymmetry  on $S^1/\mathbb{Z}_2$   can be realized
as follows:
\begin{equation}
  S_{new} =  S_{bulk}  + S_{brane}\,,\qquad  S_{bulk}  = S_{Romans}(m\rightarrow M(x))+ S_A= S_{BGPT}+{\rm fermions} \,.
\label{10newlagr}
\end{equation}
Here $S_{Romans}(m\rightarrow M(x))$ is the Romans action \cite{Romans}
where the  constant mass $m$ is replaced by a field $M(x)$ and the 9-form
field is added in $S_A= \int  \epsilon^{(10)}  dA_9 M$ so that the total
bosonic bulk action is that given by Bergshoeff, Green, Papadopoulos
and Townsend in \cite{EricAndFriends} (we have not given the fermion terms
which can be taken from the Romans action in \cite{Romans}). Note that a
slightly different version of the bosonic bulk plus brane action,
including a kinetic term for the 9-form potential and a brane kinetic
term for the BI vector fields, has been given in
\cite{Polchinski:1996df}\footnote{We obtain a kinetic term for the 9-form
potential if we eliminate $M$ as auxiliary field, rather than using the
field equation for the 9-form.}.

The brane action has to be added to the bulk action so that the on-shell
value of the $M$-field   is  piecewise constant. In particular, for an
$S^1/\mathbb{Z}_2$  orbifold, the value of $M$  should  change sign across
the wall. The parity assignments of the fields follow straightforwardly
by reducing the $D=11$ parity assignments under $y \rightarrow -y$, where
$y$ refers to the direction between the ``end of the world'' branes of
Ho\u{r}ava and Witten, over a direction different from $y$, i.e.~a
direction inside the branes.

Without the brane action, the on-shell field $M$ is the same constant
everywhere and this prevents us from getting the harmonic functions
depending on moduli $|y|$. The brane actions for the 8-brane are expected
to contain the following terms
\begin{equation}
\Purple{  \mathcal{L}_{brane}\sim -2 m \,\left( \delta (y)-\delta
(y-\tilde y )\right)
  \left( e_{(9)}e^{5\sigma \over 4}  +\ft1{9!} \varepsilon ^{\muv_1... \muv_9 }A_{\muv_1... \muv_9 }\right)}\,,
\label{8brane}
\end{equation}
where $\sigma$ is the Romans scalar which is related to the standard
dilaton $\phi$ via $\sigma = - 4/5\phi$ such that the brane tension is
proportional to $1/g_s = e^{-<\phi>}$. The coordinate $y$ refers to the
direction connecting the two branes. The modified supersymmetry
transformations in the presence of the supersymmetry singlet field $M$ and
of the 9-form are expected to exist and the bulk and the brane actions
will be supersymmetry invariant. Note that the brane action only contains
bosonic terms. It is supersymmetric due to the orbifold condition
$\psi_\muv (y) = \gamma_y \psi_\muv (-y)$.

The new physics due to the presence of the brane action will show up via
the  $M$-field equation. The bosonic part of this  equation following
from the bulk action  is given in \cite{EricAndFriends} and reads
\begin{eqnarray}
\hat{F} \equiv \epsilon \hat{F}_{10} = M(x)e K(B)
 \,. \label{feGonshell2}
\end{eqnarray}
Here $K$ is a polynomial in terms of the massive 2-form field $B$ and
$\hat{F}_{10}$ is the covariant curvature of $A_9$. With account of the
brane action (\ref{8brane}) added according to our rule of consistent
supersymmetry in the singular spaces, the on-shell value of the $M$-field
will be proportional to the step function:
\begin{eqnarray}
\hat{F}= m \varepsilon (y) e K(B)
 \,. \label{feGonshell3}
\end{eqnarray}
Clearly the flux is changing sign when passing through the brane,
according to the field equations of the theory.

One would hope that the improved supersymmetric formulation of the
8-brane of string theory in the framework of 10D supergravity in a
singular space will shed some light on the M-theory and its 9-brane.
Since 11D supergravity does not admit a cosmological constant, one may in
this way uplift from 10D to 11D some important information on extended
objects of string theory, which is difficult to understand  from the 11D
massless supergravity point of view.

A related general problem to which the construction of this paper may be
useful is to clarify the appearance of chiral  fermions on the brane in
the context of supersymmetric theory with extended supersymmetry in the
bulk.

\medskip
\section*{Acknowledgments.}

\noindent We are grateful to A. Ceresole, G. Gibbons, A. Giveon, S.
Gubser, S. Ferrara, N. Lambert, W. Lerche, A. Linde, J. Louis, A. Lukas,
D. Marolf, P. Mayr, K. Stelle, B. Ovrut, W. Unrhu and D. Waldram for
interesting and useful discussions and to the Theory Division of CERN
where  this work was initiated and to large extent performed. This work
was supported by the European Commission TMR programme ERBFMRX-CT96-0045,
in which E. Bergshoeff is associated with Utrecht university. The work of
R.K  was supported by NSF grant PHY-9870115.
\newpage
\appendix
\section{Notations}\label{app:notations}
The metric is $(-++++)$, and $[ab]$ denotes antisymmetrization with total
weight~1, thus $\ft12(ab -ba)$. We use the following indices
\begin{eqnarray}
 \mu  & 0,\ldots ,3,5 & \mbox{local spacetime}     \nonumber\\
 \muv & 0,\ldots ,3 & \mbox{4-d local spacetime} \nonumber\\
 a & 0,\ldots ,3,5 & \mbox{tangent spacetime}     \nonumber\\
 m & 0,\ldots ,3 & \mbox{tangent spacetime in 4 dimensions}\nonumber\\
  i  & 1,2 & SU(2) \nonumber\\
 I & 0,\ldots ,n & \mbox{vectors} \nonumber\\
 x& 1,\ldots ,n & \mbox{scalars in vector multiplets.}
\label{indices}
\end{eqnarray}
For the curvatures and connections, we use the conventions:
\begin{eqnarray}
\omega_\mu^{ab} &=& 2 e^{\nu[a} \partial_{[\mu} e_{\nu]}{}^{b]} -
e^{\nu[a}e^{b]\sigma} e_{\mu c} \partial_\nu e_\sigma{}^c\,,\nonumber\\
R_{\mu\nu}{}^{ab} &=& 2 \partial_{[\mu} \omega_{\nu]}^{ab} + 2
\omega_{[\mu}^{ac} \omega_{\nu]}{}_c{}^b\, ,\nonumber\\ R_{\mu\nu}&=&
R_{\mu\rho}{}^{ba}e_b{}^\rho e_{\nu a}\,, \qquad R = g^{\mu\nu}
R_{\mu\nu}\, ,  \nonumber\\
R^\mu{}_{\nu\rho\sigma}&=&R_{\rho\sigma}{}^{ab}e_a^\mu e_{\nu b}=
2\partial_{[\rho}\Gamma^\mu_{\sigma]\nu} +2\Gamma^\mu_{\tau[\rho}
\Gamma^\tau_{\sigma]\nu} \,.
\end{eqnarray}
For the metric (\ref{metric}), the only non-zero components of the spin
connection are
\begin{equation}
  \omega _\muv ^{m5}= a'\delta_\muv ^m\,.
 \label{warpedomega}
\end{equation}
Here ${}'$ means  $\partial \over \partial x^5$. The non-zero curvature
components, Ricci tensor and scalar curvature are
\begin{eqnarray}
&&  R_{5\muv}{}^{m5}=a''\delta_\muv ^m\,,\qquad
R_{\muv\nuv}{}^{mn}=-2a'^2\delta_\muv ^{[m}\delta _\nuv^{n]}\,,
\nonumber\\
&&  R_{55}=4a^{-1}a''\,,\qquad R_{\muv \nuv}= (aa''+3a'^2)\eta_{\muv
  \nuv}\,,\qquad  R= 8a^{-1}a''+12 a^{-2}a'^2\,, \nonumber\\
 && eR= -12 a^2 a'^2 +8\left(a'a^3 \right) '\,.
 \label{warpedeR}
\end{eqnarray}

The Levi--Civita tensor is real, and
\begin{equation}
  \varepsilon _{abcde}\varepsilon ^{abcde}=-5!\,,\qquad
  \varepsilon ^{\mu \nu \rho \sigma\tau }=ee^\mu _ae^\nu _b\cdots e^\tau _e\varepsilon
^{abcde}\,.
 \label{LeviCivita}
\end{equation}
The gamma matrices are related by
\begin{equation}
  \gamma ^{abcde}=\rmi\varepsilon^{abcde}\,.
\label{gammaepsilon}
\end{equation}
$SU(2)$ indices are raised and lowered with $\varepsilon _{ij}$,
where $ \varepsilon _{12}=\varepsilon ^{12}=1$, in NW--SE convention:
\begin{equation}
  X^i=\varepsilon ^{ij}X_j\,,\qquad X_i=X^j\varepsilon _{ji}\,.
\label{NWSEconv}
\end{equation}

Spinor indices are omitted. The charge
conjugation ${\cal C}$ and  ${\cal C}\gamma _a$ are antisymmetric. ${\cal
C}$ is unitary and $\gamma _a$ is hermitian apart from the timelike one,
that is antihermitian. The bar is the Majorana bar:
\begin{equation}
  \bar \lambda ^i =(\lambda ^i) ^T {\cal C}\,.
\label{barlambda}
\end{equation}
Define the charge conjugation operation on spinors as
\begin{equation}
(  \lambda ^i)^C\equiv \alpha^{-1}B^{-1}\varepsilon ^{ij}(\lambda
^j)^*\,, \qquad \bar \lambda ^{iC}\equiv \overline {(  \lambda ^i)^C}=
\alpha ^{-1}\left( \bar \lambda{}^k\right) ^*B\varepsilon ^{ki}\,,
 \label{defCspinors}
\end{equation}
where $B={\cal C}\gamma _0$, and $\alpha $ is arbitrary $\pm 1$ when you
use the convention that complex conjugation does not interchange the order
of spinors, or $\pm\rmi$ when complex conjugation does interchange the
order of spinors. Symplectic Majorana spinors satisfy $\lambda =\lambda
^C$. Charge conjugation acts on gamma matrices as $(\gamma _a)^C=-\gamma
_a$, does not change the order of matrices, and works on matrices in
$SU(2)$ space as $M^C=\sigma _2 M^*\sigma _2$. Complex conjugation can
then be replaced by charge conjugation, if for every bispinor one inserts
a factor $-1$. Then e.g. the expression
\begin{equation}
  \bar \lambda ^i \gamma ^\mu \partial _\mu \lambda _i
\label{realexpr}
\end{equation}
is real for symplectic Majorana spinors. For more details, see e.g.
\cite{Tools}.
\section{Supergravity in 5 dimensions with vector multiplets}
\label{app:5dSGVM} We present here 5-dimensional supergravity with vector
multiplets, which we will use in the main text. To put it in perspective,
we repeat that the pure supergravity was constructed in
\cite{Cremmer,Chamseddine:1980sp}. The coupling with vector multiplets
was obtained in \cite{GST84}, and with gauging of the vectors in
\cite{GST85}. This was extended to tensor multiplets in
\cite{GunZagGaugMET}, and an action coupled to an arbitrary quaternionic
manifold is obtained in \cite{AnnaGianguido}.

We consider the coupling of abelian vector multiplets to supergravity. The
fields are the f\"{u}nfbein, the gravitino, the vectors, scalars and gauginos:
\begin{equation}
  e_\mu ^a\,,\ \psi _\mu ^i\,,\ A_\mu ^I\,,\ \varphi ^{x}\,,\
  \lambda ^{ix}\,.
 \label{fields}
\end{equation}
The vector multiplets couplings are determined by the symmetric real constant
tensor $C_{IJK}$. The scalars appear in the functions\footnote{These functions are
arbitrary up to
non-degeneracy conditions that the $(n+1)\times (n+1)$ matrix $(h^I,
h^I_x)$ (see definition below) should be invertible.}
$h^{I}(\varphi)$, satisfying
\begin{equation}
  h^{I}(\varphi )h^{J}(\varphi)h^{K}(\varphi )   C_{IJK}=1\,,
\label{hhhC1}
\end{equation}
and define the quantities
\begin{eqnarray}
  &&h_I(\varphi )\equiv C_{IJK}h^J(\varphi )h^K(\varphi )=a_{IJ}h^J \,,\qquad
  a_{IJ}\equiv -2C_{IJK}h^K+3h_Ih_J\,,\nonumber\\
&&  h^I_x\equiv -\sqrt{\ft32}h^I_{,x}(\varphi )\,,\qquad
  h_{Ix}\equiv a_{IJ}h^J_x=\sqrt{\ft32} h_{I,x}(\varphi )\,, \nonumber\\
&&  g_{xy}\equiv h_x^Ih_y^J a_{IJ}=-2h_x^Ih_y^JC_{IJK}h^K\,, \qquad
 T_{xyz}  \equiv   C_{IJK}h^I_xh^J_yh_z^K\,,
  \label{defquantVM}
\end{eqnarray}
with $_{,x}$ an ordinary derivative with respect to $\varphi ^x$. The
$I$-type indices are lowered or raised with $a_{IJ}$ or its inverse,
which we assume to exist, and the same holds for the metric $g_{xy}$ used
for the indices on scalars and gauginos.

Many identities have been derived in previous papers. The most
useful ones for us are
\begin{eqnarray}
&& h_Ih^I_x=0 \,, \qquad  h_Ih_J+h_I^xh_{Jx}=a_{IJ}\,.\nonumber\\
&& h_{Ix;y}\equiv h_{Ix,y}-\Gamma _{xy}^zh_{Iz}  = \sqrt{\ft23}\left( h_I
g_{xy}+T_{xyz}h^z_I\right)  \,,
 \label{usefulVM}
\end{eqnarray}
with the connection $\Gamma_{xy}^z$ as usual defined such that
$g_{xy;z}=0$.

We consider just one gauged $U(1)$, whose coupling constant is called
$g$, and whose gauge vector is a linear combination of the vectors,
\begin{equation}
  A_\mu ^{(R)}\equiv V_I A_\mu ^I\,,
 \label{AR}
\end{equation}
where $V_I$ are real constants. It gauges a direction in $SU(2)$
space determined by a matrix
 (with real $q_1$, $q_2$ and $q_3$):
\begin{eqnarray}
&&  Q_i{}^j=\rmi(q_1\sigma _1+q_2\sigma _2+q_3\sigma _3)\,,\qquad
 (q_1)^2+(q_2)^2+(q_3)^2=1\,,\nonumber\\
&& Q_{ij}=-\rmi q_1\sigma _3-q_2\unity +\rmi\sigma _1 q_3\,,\qquad
Q^{ij}=\rmi q_1\sigma _3-q_2\unity -\rmi\sigma _1 q_3\,,\nonumber\\
&&  Q_{ij}Q^{jk}=\delta _i^k\,,\qquad Q_i{}^jQ_j{}^k=-\delta _i^k\,.
 \label{Qreal}
\end{eqnarray}
The equations of \cite{GST85} are obtained by choosing $q_1=q_3=0$,
$q_2=1$. They, and \cite{AnnaGianguido}, introduce also other functions
that are useful when one considers non-abelian gauging:
\begin{equation}
P_{Iij}=V_IQ_{ij}\,,\qquad   P_{ij}=h^IV_I Q_{ij}\,,\qquad
P^x_{ij}=h^{xI}V_IQ_{ij}\,.
 \label{defPmatrices}
\end{equation}
We will use the quantities
\begin{equation}
 W  =  \sqrt{\ft23}h^I V_I\,,
 \label{defnW}
\end{equation}
with derivative
\begin{equation}
  W_{,x}\equiv \frac{\partial }{\partial \varphi ^x}W=-\frac 23
  h_x^IV_I\,.
 \label{Wx}
\end{equation}

The action is then $S_{GST}=\int \rmd^5x\, {\cal L}_{GST}$, with
\begin{eqnarray}\label{LagrangeVectorExpl}
e^{-1}\mathcal{L}_{GST}&=& -\ft{1}{2}R(\omega)-\ft{1}{2}
{\bar{\psi}}_{\mu}^{i}\gamma^{\mu\nu\rho}\left(
 \nabla_{\nu}(\omega)\psi_{\rho i} +gA^{(R)}_\nu Q _{ij}\psi^j _\rho\right)
-\ft{1}{4}a_{IJ} F_{\mu\nu}^{I} F^{J\mu\nu}\nonumber\\
 && +
\ft{1}{6\sqrt{6}}e^{-1}\varepsilon^{\mu\nu\rho\sigma\lambda}
C_{IJK} F^I_{\mu\nu}F^J_{\rho\sigma}A^K_{\lambda}
\nonumber\\
&&-\ft12g_{xy} \partial _\mu \varphi
^x \partial ^\mu \varphi
^y-V(x)\nonumber\\ &&
 -\ft{1}{2}\overline{\lambda }{}^i_x\gamma ^\mu\left( \nabla _\mu
(\omega )\lambda ^x_i+\Gamma ^x_{yz}(\partial _\mu \varphi ^y)
\lambda ^z_i  +g A^{(R)}_\mu Q _{ij}\lambda ^{jx}\right)
\nonumber\\
  &&-\rmi\ft{\sqrt{6}}{16}\left[{\bar{\psi}}_{\mu}^{i}
\gamma^{\mu\nu\rho\sigma} \psi_{\nu i}F_{\rho\sigma}^Ih_I +2{\bar{\psi}}^{\mu
i} \psi_{i}^{\nu}F_{\mu\nu}^Ih_I\right]
\nonumber\\
 &&  +\ft14 \overline{\lambda }{}^{ix}\gamma ^\mu \gamma ^{\nu \rho }\psi _{i\mu }F_{\nu \rho
 }^Ih_{Ix} -\ft\rmi 2 \overline{\lambda }{}^i_x\gamma ^\mu \gamma ^\nu
\psi _{i\mu } \partial _\nu \varphi ^x
 \nonumber\\
 &&
  +\ft\rmi 4\sqrt{\ft23}(\ft14 g_{xy}h_I+T_{xyz}h^z_I)
 \overline{\lambda }{}^{ix}\gamma ^{\mu \nu }\lambda _i^yF_{\mu \nu }^ I
 \nonumber\\
 && +g Q_{ij}\left[ \rmi\overline{\lambda }{}^{ix}\lambda ^{jy}
\left(- \ft{1}{4}g_{xy}W+\sqrt{\ft32} T_{xyz}W^{,z}\right)
-\ft32\overline{\lambda }{}^i_x W^{,x}\gamma ^\mu \psi ^j_\mu
+\rmi\ft3{4}\overline{\psi}{}^i_\mu \gamma ^{\mu \nu }
\psi _\nu ^j W \right] \nonumber\\
&& + \mbox{4-fermion terms.}
\end{eqnarray}
where ($W^{,x}\equiv g^{xy}W_{,y}$)
\begin{eqnarray}
 \nabla _\mu  & = & \partial _\mu +\ft14 \omega _{\mu ,ab}
  \gamma ^{ab}\,, \qquad
  F_{\mu \nu }^I=2\partial _{[\mu }A_{\nu ]}^I \,,\nonumber\\
 V(x) & = &-4g^{2}V_IV_JC^{IJK}h_K=g^{2}\left(-6W^2+\ft92W^{,x}W_{,x}\right)
 \,.
 \label{defnV}
\end{eqnarray}
The action is invariant under the supersymmetry transformations
\begin{eqnarray}
\delta (\epsilon )e^a_\mu  &=& \ft{1}{2} \bar{\epsilon}{}^i
\gamma^a \psi_{\mu i}\,,\nonumber\\
\delta (\epsilon )\psi_{\mu i} &=&
\partial  _\mu  \epsilon _i+ \ft14
\gamma^{cd} \hat{\omega}_{\mu , cd}\epsilon _i  + gA^{(R)}_\mu
Q_{ij}\epsilon  ^j  + \frac{\rmi}{4\sqrt{6}} \left( \gamma_{\mu \nu \rho
} - 4 g_{\mu \nu } \gamma_\rho  \right)\epsilon_i h_I \widehat{F}{}^{\nu
\rho \,I}\nonumber\\
&& - \ft{1}{12} \gamma _{\mu \nu }\epsilon ^j\overline{\lambda
}{}^x_i\gamma ^\nu \lambda _{jx}+\ft{1}{48}\gamma_{\mu \nu \rho }\epsilon
^j\overline{\lambda }{}_i^x\gamma ^{\nu \rho }\lambda
_{jx}+\ft{1}{6}\epsilon ^j\overline{\lambda }{}_i^x\gamma _\mu \lambda
_{jx}-\ft{1}{12}\gamma ^\nu \epsilon ^j\overline{\lambda }{}^x_i\gamma
_{\mu \nu }\lambda _{jx}\nonumber\\
&&+\rmi\ft12  g \, \gamma_\mu \epsilon^j WQ_{ij}\,,
\nonumber\\
\delta(\epsilon )A^I_\mu  &=& -\frac{1}{2}\overline{\epsilon }^i\gamma
_\mu \lambda _i^xh_x^I+\frac{\rmi\sqrt{6}}{4}h^I \overline{\psi}_\mu
^i\epsilon_i \,, \nonumber\\
\delta(\epsilon )\varphi ^x&=&\frac{\rmi}{2}\overline{\epsilon }^i\lambda _i^x\,,\nonumber\\
\delta(\epsilon )\lambda _i^x&=&-\frac{\rmi}{2}\widehat{\dsl}\varphi
^x\epsilon _i +\frac{1}{4}h_I^x\gamma ^{\mu \nu }\epsilon
_i\widehat{F}_{\mu \nu }^I
-g\ft32 \epsilon ^jW^{,x}Q_{ij}\nonumber\\
&& -\frac{\rmi}{2}h_I^xh_{Iy,z}\lambda _i^y\overline{\epsilon }^j\lambda
_j^z +\frac{\rmi}{\sqrt{6}}\epsilon ^j\overline{\lambda }{}_i^y\lambda
_j^z T_{xyz}
\,,\label{susy5dV}
\end{eqnarray}
where
\begin{eqnarray}
  \hat{\omega }_{\mu ab}&=&\omega _{\mu ab}
  -\ft14\left(\bar \psi ^i_b\gamma _\mu \psi _{ai}
  +2\bar \psi^i_\mu \gamma _{[b}\psi _{a]i}\right) \,, \nonumber\\
   \widehat{F}_{\mu \nu }^I &=& 2\partial
_{[\mu }A_{\nu ]}^I +h_x^I\,\overline{\psi}{}^i_{[\mu }\gamma_{\nu ]}
\lambda _i^x+\ft14
\rmi\sqrt{6}\; h^I\,\overline{\psi}^i_{[\mu } \psi_{\nu ]i}\,,\nonumber\\
\widehat{\partial} _\mu \varphi ^x & = & \partial _\mu \varphi
^x-\ft{\rmi}{2} \overline{\psi}{}_\mu ^j\lambda _j^x\,.
 \label{defTransfVM}
\end{eqnarray}

\section{Previous work on supersymmetry in singular spaces}\label{app:previous}

The first idea to have two supersymmetric domain walls relating the
11-dimensional physics with the 10-dimensional one, was suggested in HW
\cite{HoravaWitten}. They made the connection between the bulk and the
branes at the quantum level, i.e. 1-loop anomalies of the bulk action are
cancelled by a classical non-invariance of the action on the brane. The
non-trivial flux that introduces the step functions into the theory, is
related to an anomaly. The realization of the supersymmetry of this
combined bulk and brane actions was successful at the order
$\kappa^{2/3}$, which was established by rather involved calculations.
However, at the order $\kappa^{4/3}$ things go out of control. Some
$\delta(0)$ terms occur. The hope was expressed that a proper quantum
M-theory treatment will lead to quantum consistency. The important aspect
of this theory was the appearance of chiral fermions on the brane.

The next step in this development was undertaken by \cite{Kellyandco},
where HW theory was compactified on a CY space and the ``5D supergravity''
with the step functions in the supersymmetry rules, was recovered. The
supersymmetry in this approach relies on the supersymmetry of HW theory
and therefore would be difficult to complete.

More recently, two new  approaches to realize the supersymmetric RS
scenario were suggested in the framework of 5D supergravity:
one in \cite{Bagger} and an alternative one in \cite{Pomarol,Polish}. The basic
difference is that in \cite{Bagger} the gauge coupling constant is even
across the wall and in \cite{Pomarol,Polish} it is taken to be odd as in HW
theory. We have shown in section~\ref{ss:susyZ2}, using a standard basis
for symplectic spinors, that there are indeed two possibilities to make
the $\mathbb{Z}_2$ symmetry on spinors commuting with the supersymmetry:
one with even  gauge coupling, when the projector in parity rules
anticommutes with the projector in supersymmetry rules, and the second
one with the odd gauge coupling, when the projector in parity rules
commutes with the projector in supersymmetry rules. So far both versions
were fine.

In the case of even $g$ \cite{Bagger}, the bulk action is simply the
action of supergravity, without step functions: therefore the standard
supersymmetry rules are valid for the bulk action and there is no
compelling reason to add the brane actions. The motivation for adding the
brane actions comes from the properties of the singular background: if one
would consider configurations with  unbroken supersymmetries for domain
walls, one would not solve Killing equations everywhere. This enforces
some particular changes in supersymmetry rules at the singularities only.
The brane actions are added to compensate the supersymmetry of the bulk
action, which is not supersymmetric after the $\delta$-functional changes
in the supersymmetry rules. We have checked that if one would start not
from pure supergravity without matter, as in \cite{Bagger}, but with
additional multiplets, the change of supersymmetry rules would depend on
the scalars in the background and the issue of the closure of the algebra
would become obscure as the scalars had to satisfy equations of motion.
The conclusion of this analysis was that it will be difficult to
generalize the approach of \cite{Bagger} in pure supergravity to more
general theories with matter multiplets.

We therefore decided to continue along the line of \cite{Pomarol,Polish}, where
the gauge coupling $g$ was  taken to be odd. This enforced us to
introduce step functions in the supersymmetry rules and as a consequence
the bulk supergravity action was not supersymmetric anymore: the brane
action of \cite{Polish} was designed to compensate the problem of the
bulk action. This was conceptually and technically correct from our
perspective. The problem was  how to make the total system complete.
Later when the supersymmetry singlet and the 4-form potential were
introduced, we were able to construct a consistent and complete
supersymmetric theory where the bulk and the brane actions are
supersymmetric. Remarkably, if we would not care about the conceptual
issues and would only try to accomplish the complete supersymmetric
theory, we would have to use the odd gauge coupling to have a consistent
implementation of the 4-form field, see section~\ref{ss:branesink}, where
this is explained.

Here we also would like to comment on a very interesting paper
\cite{Tomas}. It does not deal with fermions, therefore it does not
address directly the issues discussed in our paper. However, it brings
some deep insights from the perspective of the brane actions: in
particular the role of the $(D-1)$-form is stressed as required for the
Wess--Zumino terms of codimension~1  branes, as well as the fact that the
bulk action has to be supplemented by the term that is quadratic in the
$D$-form flux. In our supersymmetric theory, we have not eliminated the
field $G$ as an auxiliary field by its equation of motion. If we would
have done it using the action in the form of (\ref{LagrangeVectorBulk})
and (\ref{SbraneA}), we would find out that the bulk action has a term
quadratic in the flux $\hat F$. We expect that our supersymmetric action
with the brane action where the worldvolume degrees of freedom are not
excited, may be further developed in the spirit of \cite{Tomas} and lead
to a deeper understanding of the total dynamical system.


\end{document}